\documentclass[11pt]{elsarticle}

\usepackage{graphics,graphicx}

\usepackage[T1]{fontenc}
\usepackage[latin1]{inputenc}
\usepackage{graphics}
\usepackage{setspace}
\usepackage{amsmath}
\makeatletter

\providecommand{\LyX}{L\kern-.1667em\lower.25em\hbox{Y}\kern-.125emX\@}

\journal{Physica A}
\begin{document}

\title{Estimate of time-scale for the current relaxation of percolative Random Resistor cum Tunneling Network model}

\author[label1]{Somnath Bhattacharya} 
\ead{somnath94347@gmail.com}
%\author[label2]{}
%\ead{}
%\email[]{lucas.goehring@ds.mpg.de}
%\thanks{}
\address[label1]{Postgraduate Department of Physics, Barasat Govt. College, 10, K. N. C. Road, Kolkata-700 124, India}
%\address[label2]{}
%\ead[label1]{soumyajyoti.biswas@ds.mpg.de}
%\ead[label2]{psphy@caluniv.ac.in}

\date{\today}

\begin{abstract}
\noindent  The Random Resistor cum Tunneling Network (RRTN) model was proposed from our group by considering an extra phenomenological 
(semi-classical) tunneling process into a classical RRN bond percolation model. We earlier reported about early-stage two 
inverse power-laws, followed by large time purely exponential tail in some of the RRTN macroscopic current relaxations. In 
this paper, we investigate on the broader perspective of current relaxation. We present here an analytical argument behind the
strong convergence (irrespective of initial voltage configuration) of the bulk current towards its steady-state, mapping the
problem into a special kind of Gauss-Seidel method. We find two phenomenological time-scales (referred as $\tau_t$ and $\tau_s$), those
emerge from the variation of macroscopic quantities during current dynamics. We show that not both, only one of them is independent. Thus there exists a {\it single} scale in time which controls the entire dynamics. 
\end{abstract}
\begin{keyword}
Percolation \sep Electrical Properties \sep Relaxation \sep Time-scale 
\end{keyword}
%\pacs{72.80.Tm, 61.43.-j, 64.60.Ak, 67.40.Fd, 78.20.Bh}

\maketitle

\section{Introduction}    

Relaxation phenomena in binary composite materials, which are made of two components with widely different generalized 
susceptibilities, draw considerable interest among the scientists and engineers, even in very recent research activities 
\cite{water,bism,PVC}. The susceptibility (being a measure of the {\it response} of a system to an appropriate external field) is approximately treated as {\it linear} upto a sufficiently small field. In general the {\it nonlinearity} in the response starts manifesting for large external drive. 
However, there are several soft condensed matter systems (e.g., binary composites) where measurable response starts only above a low 
{\it threshold} value of the external field. Above this threshold, these systems show nonlinear response because then the generalised susceptibility is seen to be a function of external field.

In studying the relaxation behaviour of an open physical system, one usually measures its appropriate response property, say, 
$\phi(t)$, as a function of time $t$, during its passage from a non-steady to a steady state.  In general, this relaxation is 
classified into two groups: (i) a purely Debye type with an exponential relaxation function, $\phi(t)= {\rm exp}(-t/\tau)$, $\tau$
 being a characteristic time-scale, called the {\it relaxation time}; or (ii) a non-Debye type where $\phi(t)$ can not be mathematically expressed by a single exponential function associated with a single $\tau$. It is either a linear superposition of several exponential functions having different time constants, or may be a sub-exponential function (as in some glassy systems), with {\it multiple} relaxation times. Even this non-Debye class of relaxation includes the cases where $\phi(t)$ is represented by a single or couple of power-laws or a logarithmic function. This class of dynamics lacks any finite time-scale or sometimes the relaxation behaviour is {\it scale-free}.

The Random Resistor cum Tunneling Network (RRTN) model was primarily proposed \cite{dc} by Sen et. al. decades ago. This
model considers an extra phenomenological (semi-classical) tunneling process into a classical RRN bond percolation model. We earlier reported \cite{sust,epl05} about intriguing early stage two inverse power-laws in time for {\it some} of the RRTN macroscopic current relaxations. One may appropriately note here that there are huge varieties of natural as well as synthesized systems, which manifest non-Debye type power-law relaxation in their appropriate responses. In our earlier papers \cite{sust,epl05}, one may find short description on some of these experiments and theoretical investigations. We mention here about two recent works on power-law relaxation to emphasize that the study in this line is still relevant. A coupled memory was introduced in case of continuous-time random walk \cite{jurle} to generate a two power-law relaxation behaviour from Debye type relaxation. Also in a subordination model of anomalous diffusion \cite{kweron} power-law relaxation behaviour appears.  However in addition to most frequent observation in case of two power-law relaxation, we could show an intriguing type of RRTN dynamics from a percolative paradigm \cite{epl05}, where the exponent of second power-law is smaller than the same for the first one, which is seen in experiments \cite{bez,bar,sangam}, but was not theoretically explored earlier than \cite{epl05}. Recently also in an experiment with worm-like micellar solution, a similar behaviour \cite{less2nd} was reported for the response quantity like mean square displacement. These discussions surely establishes the richness of the RRTN model in explaining some versatile dynamical behaviours of soft percolative systems including some of the recent findings. 

However in this present paper, we investigate on a broader perspective of current relaxation in the RRTN model. One may appropriately note here that all RRTN current dynamics show two regimes (i) an early-stage non-exponential regime, a signature of nonlinear cooperation among the sub-systems, and (ii) a latter-stage purely exponential regime with a single $\tau$, usually seen in linear systems for the entire relaxation dynamics.  One may note here that the sequence of appearance of regimes is opposite to that in ref. \cite{jurle}. Broadly speaking, only for few RRTN dynamics, the early-stage regime shows two-power laws, some show a single power-law during initial regime. Even for few cases the non-Debye regime can not at all be expressed via any power-law (i.e., one or two for at least one decade). More interestingly irrespective of the early-stage functional form, {\it every} RRTN bulk current relaxation has the exponential tail after the cross-over from some non-exponential regime.  Here we address on the phenomenological reason behind the origin of these two regimes in the RRTN current dynamics.  Our study will enable us to identify the extent of these regimes in terms of two characteristic scales (in time) as system parameters. Moreover we may also indicate here that these scales do not evolve independently, so that there is a single time-scale which dictates the entire dynamics. Another interesting feature of RRTN current dynamics we observed since ref. \cite{sust}, that for each particular RRTN lattice the bulk current converges to a robust steady-state value $I_{steady}$. This happens for any kind of initial guess for microscopic voltage configuration at each node of the RRTN lattice. This was referred before as {\it intrinsic memory} \cite{aks-memo}. In this paper, we find also an analytical explanation for this strong convergence of the bulk current.

\section{The RRTN model}

For the investigation on the basic physics behind the transport properties of binary percolative composites, Sen and Kargupta constructed a lattice based bond percolation model \cite{dc}, and named it as {\it Random Resistor cum Tunneling Network} (RRTN) model.  
In this RRTN model, in addition to the randomly placed ohmic bonds (named as o-bonds, each of which shows linear response under non-zero applied voltage with conductance $g_o$) in the Random Resistor Network (RRN), some {\it tunneling} bonds (t-bond) were introduced between two nearest neighbour ($nn$) ohmic bonds, i.e., when two o-bonds are separated by one lattice constant only. This tunneling is actually claimed to be {\it semi-quantum} or semi-classical in the sense that here no quantum mechanical phase-information of the charge carriers appears in the model.  A t-bond was considered to be insulating if the magnitude of the microscopic voltage difference across it ($v$) is less than a fixed voltage threshold ($v_g$, identical for all the t-bonds). It conducts linearly a current if $|v| \ge v_g$.  In addition to that, we have introduced \cite{epl05} an additional contribution of {\it displacement current} contribution for each inactive t-bond (i.e., when $|v| < v_g$), which behaves as a dielectric material. This new consideration never contributes in the steady state current, so that the earlier results on this model \cite{dc,ac,break,vrh} remain same even with this change. The details of the algorithm on current dynamics will be discussed in the next section. Indeed, the phenomenological parameters like $p$ (concentration of o-bonds in the RRN) and $v_g$ ensure that both the {\it disorder} and the {\it coulomb interaction} is in-built in this model.

%%%%%%%%%%%%%%%%%%%%%%%%%%%%%%%%%%%%%%%%%%%%%%%%%%%%%%%%%%%%%%%%%%%%%%%%%%%%%%%%%%%%%%%%%%%%%%%%%%%%%%%%%%%%%%%%%%%%%%%%%%%%%%%%%%%%%%%%%
\begin{figure}[htb]
%\onefigure{conf.ps}
\resizebox*{7cm}{7cm}{\rotatebox{270}{\includegraphics{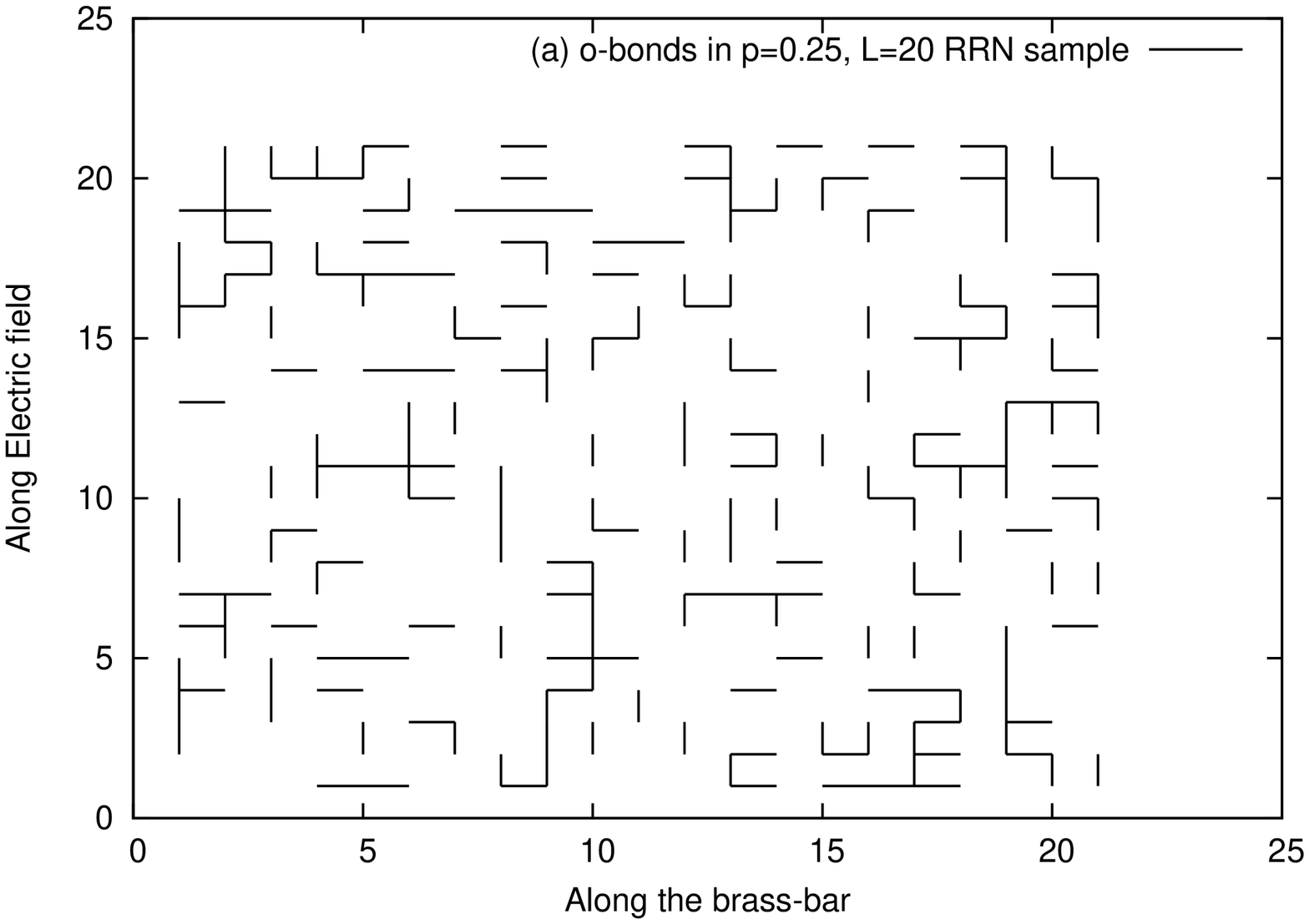}}}
\resizebox*{7cm}{7cm}{\rotatebox{270}{\includegraphics{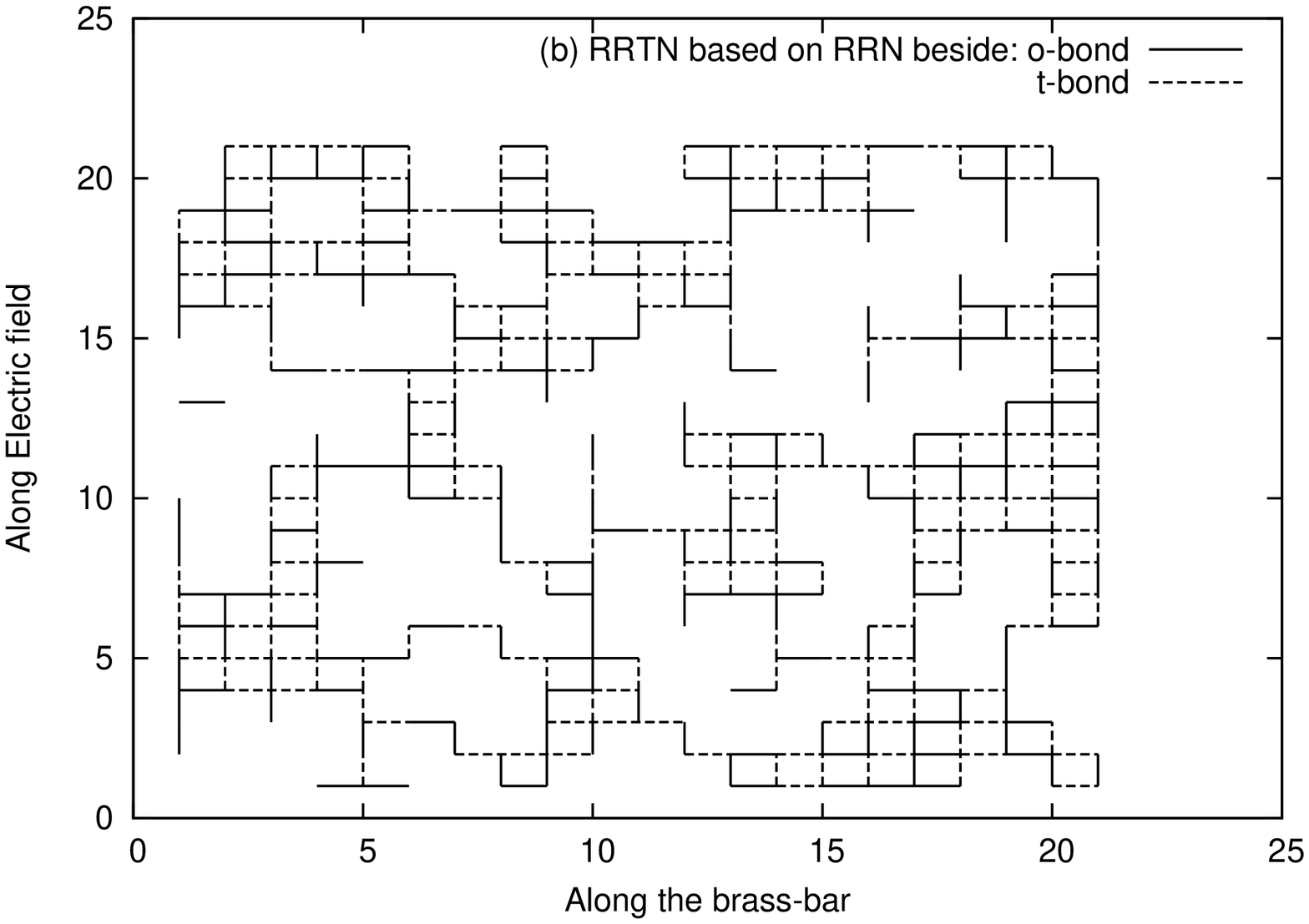}}}
\caption{(a) A representative non-percolating configuration of random metallic o-bonds in a $L=20$, $p=0.25$ RRN sample on a square lattice.  The insulating bonds are kept blank. (b) The {\it maximal} (with fully statistically correlated tunneling t-bonds) RRTN 
configuration, obtained from given RRN by placing the t-bonds (dashed lines) between any two nearest neighbour o-bonds.
Each t-bond gets activated when the potential difference across it exceeds a pre-assigned microscopic voltage threshold, 
\(v_g\).  With all the allowed t-bonds active, one may note that this maximal RRTN percolates (i.e., is a percolating RRTN).}
\label{f.1}
\end{figure}
%%%%%%%%%%%%%%%%%%%%%%%%%%%%%%%%%%%%%%%%%%%%%%%%%%%%%%%%%%%%%%%%%%%%%%%%%%%%%%%%%%%%%%%%%%%%%%%%%%%%%%%%%%%%%%%%%%%%%%%%%%%%%%%%%%%%%%%%%%
The appearance of the t-bonds in this {\it perfectly correlated} (i.e., deterministic), fashion is the origin of a very low percolation threshold in the RRTN model. In this respect, one may note that the percolation threshold for the RRN (square lattice) was $p_c = 0.5$ in the thermodynamic limit, whereas under finite-size scaling analysis for {\it maximal} RRTN (i.e., when all t-bonds are active), one finds $p_{ct} = 0.18$. For example, in the fig.\ref{f.1}(a), the RRN with $p=0.25$ does not percolate, but the maximal RRTN generated from there percolates as shown in the fig.\ref{f.1}(b). Further, because of the finite threshold of the t-bonds ($v_g$), each RRTN which is generated from a non-percolating RRN starts percolating above a macroscopic voltage threshold (say, $V_g$). As its consequence the bulk dc conductance $G(V)$ follows a strongly {\it nonlinear} S-shaped curve with increase of external dc voltage \cite{dc}. Other than the success in understanding various aspects of nonlinear dc \cite{dc} and ac responses \cite{ac} in composites with ultra-low percolation threshold, the earlier studies on RRTN model were also been quite useful for some interesting aspects of breakdown phenomena \cite{break} and understanding some very unusual aspects observed in low-temperature variable range hopping conduction \cite{vrh,ictp}. Thus, even though, time enters in an implicit fashion in some of the above studies, an explicit characterization of the relaxation dynamics in the RRTN model was considered necessary and some of its successes in the line of early-stage two-power law relaxation were reported in \cite{sust,epl05}.

\section{Lattice Kirchhoff's dynamics in the RRTN model}

As we discussed in the previous section, a t-bond with a microscopic voltage as $|v| < v_g$ behaves like a dielectric material between two
metals (o-bonds) and the resulting charging effect gives rise to a {\it displacement current} ($C\frac{dv(t)}{dt}$, where $C$ is the capacitance). However for $|v| \ge v_g$, a t-bond gives rise an ohmic current with a voltage-independent microscopic conductance (say $g_t$). For our numerical calculations, we use the values of the microscopic conductance for o-bond as $g_o=1.0$, for t-bonds $g_t = 10^{-2}$ with $v_g=0.5$, and $C=10^{-5}$ for the t-bonds (all in some arbitrary units). In our numerical study, we apply an uniform voltage across RRTNs of different system sizes ($L$) and ohmic bond concentrations ($p$).  We study the evolution of the current in a RRTN starting from the switching on state of external voltage ($V$) until the current approaches its {\it asymptotic} steady-state value. To do this, we intend to follow the current conservation (Kirchhoff's laws) locally at each node of the lattice.  The aim is to study the achievement of a {\it global current conservation} as an outcome of the {\it local current conservation} (hence, the dynamics).  A discrete, {\it scaled} time unit has been chosen as one completes scan through each site of the lattice.  This local conservation or the equation of continuity reads as, 

\begin{equation}
%\begin{displaymath}
\label{eq.1}
\sum\limits_{nn} i_{nn}(t) = 0, \hskip 2.0cm  {\forall t}.  
\end{equation}
%\end{displaymath}

\noindent
Here the {\it sum} has been taken over currents $i_{nn}(t)$ through
various types of nearest neighbour ($nn$) microscopic bonds
around any node/site of the lattice.  For the case of a square lattice,
one considers the four $nn$'s around a node inside the bulk (three and
two $nn$'s respectively at any boundary or a corner).  If eq.~(\ref{eq.1})
were true simultaneously for each site of the lattice, then the global
conservation (the steady state) for the entire network would automatically
be achieved.  As we need to start with an initial (arbitrary) microscopic
voltage distribution, the eq.~(\ref{eq.1}) would not
hold for all the sites of the lattice.  Some correction term would be
required at each site and this requirement leads to the following time
evolution algorithm which we call as the {\it lattice Kirchhoff's dynamics}:

\begin{equation}
{v(j,k,t+1)} = {v(j,k,t)} + \frac{\sum\limits_{nn} i_{nn}(t)} {\sum\limits_{nn} g_{nn}},
\label{update}
\end{equation}

\noindent
where $g_{nn}$ are the various microscopic conductances of the $nn$
bonds around the node $(j,k)$ and $v(j,k,t)$ is the microscopic voltage for node $(j,k)$ at time $t$.
Then we numerically solve a set of coupled {\it difference} equations on the lattice. For this numerical work, 
we have followed the standard {\it Gauss-Seidel} (GS) procedure for solution of coupled algebraic equations. The
iterative update continues till all the microscopic voltages $v(i,j)$ (i.e., the roots of the GS algorithm) converge to 
their steady values. The move towards a macroscopic steady-state implies that the difference of currents through the first 
and the last layers tends to zero as a function of time.  In practice, the system is considered to have reached its steady state when this 
difference decreases to a pre-assigned smallness. For the present study, we have considered the initial voltage configuration 
by assigning some random fluctuation in the graded voltage configuration at each layer as we followed in ref. \cite{sust,epl05}. 
By {\it graded} values, we mean the steady $v(i,j)$ for each layer of an ordered ($p=1.0$) square bond-percolation lattice. We refer this kind of initial guess as `arbitrary voltage configuration'.

\section{Results and Discussions}

In this paper, we investigated on some of the general characteristics of the bulk current relaxation in the RRTN network. Firstly we shall address on the strong convergence of the RRTN dynamics from the perspective of GS iterative rule. Analytically we will cite the reason behind the robust steady value for every RRTN bulk current relaxation. Secondly we will search for the phenomenological time-scales (we call them as $\tau_t$ and $\tau_s$.) in-built in the bulk current dynamics. The time-scales will qualitatively indicate the extent of the corresponding two different (i.e., non-exponential and exponential) regimes in the RRTN current relaxation. Finally we will investigate on the statistics of these time-scales for several (say, $5000$) RRTN samples of same macro-state $(L,p,V)$. From there we observe the strong correlation between $\tau_t$ and $\tau_s$, which enables us to identify a {\it single} time-scale (e.g., their ratio i.e., $r=\tau_s/\tau_t$), present in the RRTN current relaxtion.

\subsection{Convergence of RRTN current relaxation}
We reported earlier \cite{sust,epl05,aks-memo} that the current relaxation process for any {\it particular} RRTN lattice [i.e., a single bond configuration for fixed $(L,p,V)$] ends to a robust steady-state bulk current value, irrespective of \textit{any} initial microscopic voltage configuration at each node of the lattice. As discussed previously, to achieve the steady microscopic voltage distribution we always follow the lattice Kirchhoff's dynamics based on a GS algorithm. From the perspective of RRTN relaxation, the microscopic voltages (e.g., denoted by $v(i,j,t)$ for $(i,j)$ node at time $t$) are the roots of the coupled algebraic equations whereas the microscopic conductance values of each bond for the maximal RRTN are the coefficients of those equations in the GS method. Thus the convergence of RRTN bulk current to a robust steady-value is analogous to attainability of convergent roots in the Gauss-Seidel algorithm. But unlike to the conventional GS iterative method with constant coefficients in numerical analysis, in our present study the coefficients are partially time-dependent.  Because the microscopic conductance of each tunneling bond depends on microscopic voltage difference across it, which may change for each consecutive iterations. To establish the convergent criterion for RRTN current dynamics, we have considered a prototype $3 \times 3$ perfect square lattice with arbitrary values of microscopic conductance for each bond. The brass-bars are set to constant dc voltages by assigning $v(1,j,t)=0~\forall j$ and $v(4,j,t)=V~\forall j$ for entire relaxation period  (i.e., $\forall t$). The microscopic voltages for other nodes (i.e., $j \neq 1$ and $j \neq 4$) are set to be some arbitrary $v(i,j,t)$ values, which are intended to be updated for each iteration (i.e., $t$) till steady bulk current values are attained. To enumerate the bonds on the square lattice, we start with first layer (i.e., between $j=1$ and $j=2$) of vertical bonds, then second layer (i.e., for $j=2$) of horizontal bonds etc., and we continue this tagging till the last layer (i.e., between $j=3$ and $j=4$) of vertical bonds. By horizontal/vertical bond, we mean the bond in the RRTN as parallel/perpendicular to the brass-bar. One notes that we have skipped the bonds attached with the brass-bars during enumeration as for them the voltages at their both ends remain unchanged during the GS iterations.  With this frame work, we write the relevant coupled algebraic difference equations. From there we frame the corresponding coefficient matrix $\cal{C}$ as follows,
$$
%\hskip -4.3cm 
\quad 
%\resizebox{\commonwidth}{!}{
\begin{bmatrix}
C_{11}      & -g_5        & 0         & 0          &-g_8    &0       & 0       & 0 \\
-g_5        & C_{22}      & -g_6      & 0          & 0      &-g_9    & 0       & 0 \\
0           &  -g_6       & C_{33}    & -g_7       & 0      &0       &-g_{10}  & 0 \\
0           & 0           & -g_7      & C_{44}     & 0      &0       &0        &-g_{11} \\
-g_8        & 0           & 0         & 0          & C_{55} &-g_{12} & 0       &0 \\
0           & -g_9        & 0         &0           &-g_{12} &C_{66}  &-g_{13}  & 0 \\
0           & 0           & -g_{10}   & 0          &0       &-g_{13} & C_{77}  & -g_{14} \\
0           & 0           &0          & -g_{11}    & 0      &0       & -g_{14} & C_{88}
\end{bmatrix}
%}
\quad
$$

where, diagonal elements of matrix $\cal{C}$ are,
\begin{eqnarray*}
C_{11} &=& g_1+g_5+g_8, \\
C_{22} &=& g_2+g_5+g_6+g_9,\\
C_{33} &=& g_3+g_6+g_7+g_{10},\\
C_{44} &=& g_4+g_7+g_{11},\\
C_{55} &=& g_8+g_{12}+g_{15},\\
C_{66} &=& g_9+g_{12}+g_{13}+g_{16},\\
C_{77} &=& g_{10}+g_{13}+g_{14}+g_{17},\\
~and~C_{88} &=& g_{11}+g_{14}+g_{18}. 
\end{eqnarray*}

One may find that this coefficient square matrix is diagonally dominant as well as symmetric. This is the necessary and sufficient condition for the convergence in a GS process \cite{numa}. This establishes an analytical argument behind the strong convergence of the RRTN current dynamics for any kind of microscopic voltage configuration. One may follow any style for enumeration of bonds in a lattice, based on that each term of the matrix elements will be renamed. But the necessary important property of the coefficient matrix will remain intact irrespective of any specific enumeration scheme for the bonds. For a disordered network like RRN, some of the $g_i$s will be zero. For a RRTN, in addition to these resistors /o-bonds, those $g_i$s which corresponds to microscopic conductances for tunneling bonds change their value in time. So the coefficient matrix varies for different iterations.  But as the above matrix with any arbitrary finite values of $g_i$s satisfies the required properties for the convergence in GS algorithm, so our argument is very much true for RRTN current relaxation.

\subsection{On search for the time-scale during relaxation}

\begin{figure}[htb]
\resizebox*{7cm}{7cm}{\rotatebox{270}{\includegraphics{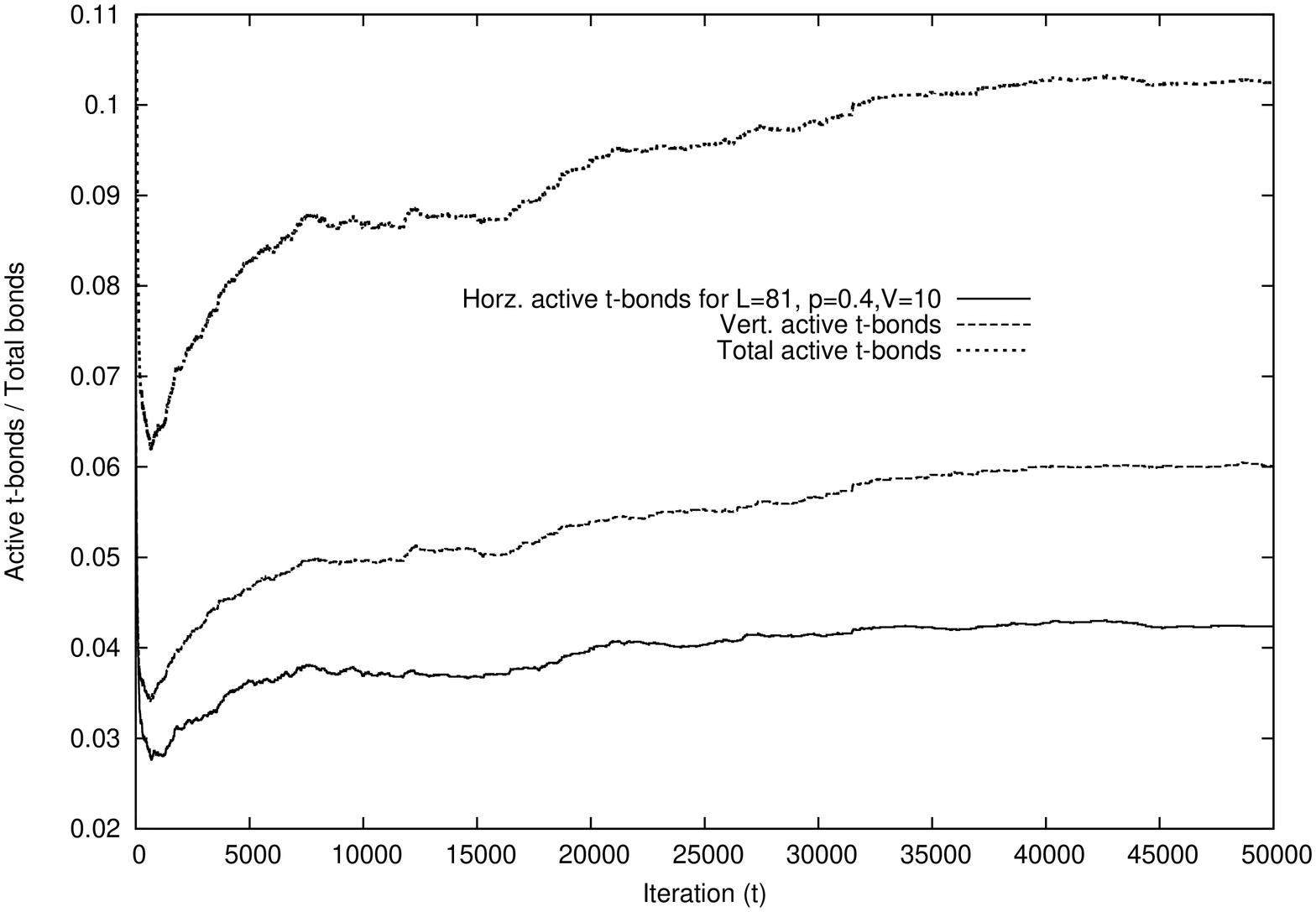}}}
\resizebox*{7cm}{7cm}{\rotatebox{270}{\includegraphics{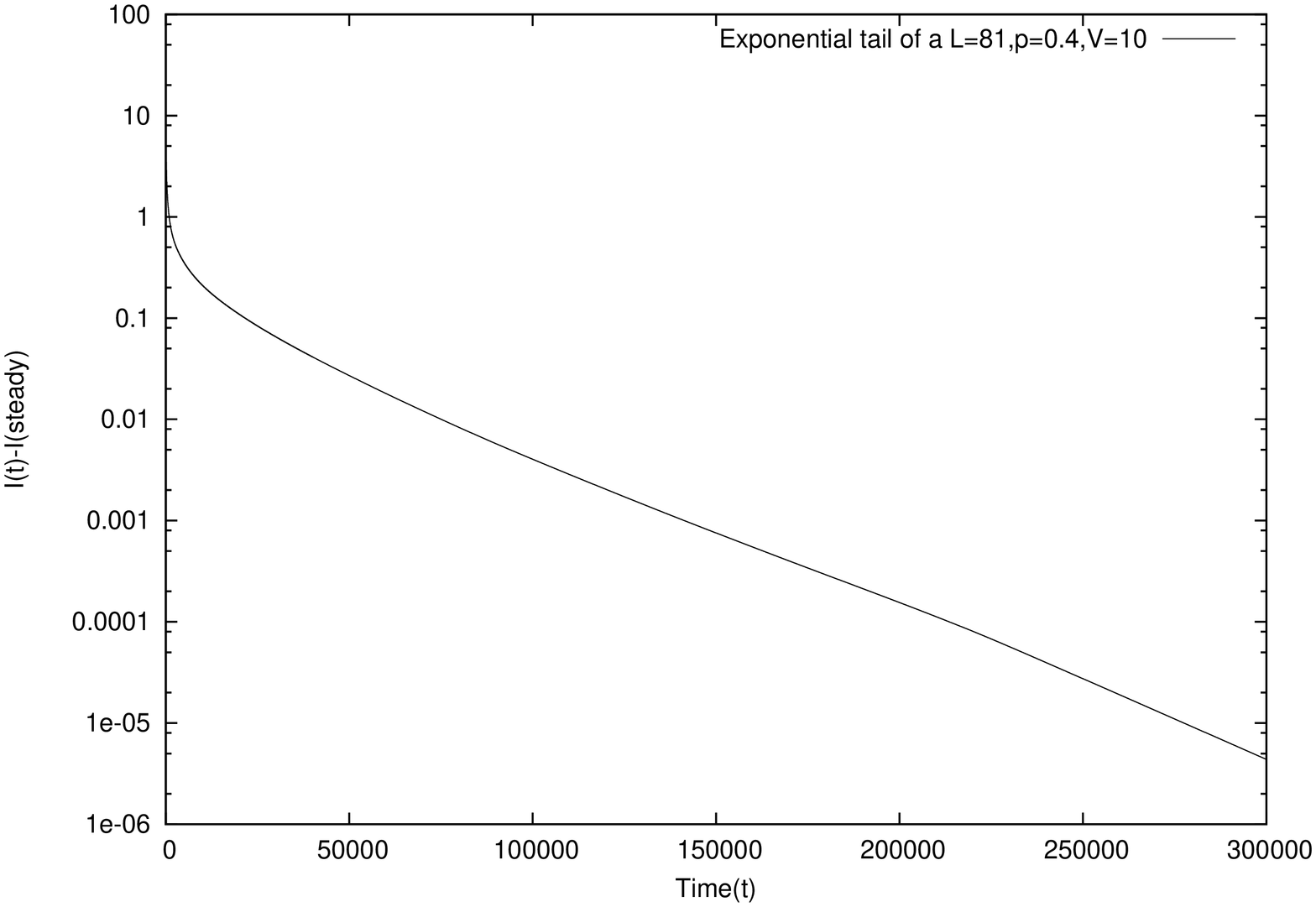}}} 
\caption{(a) A variation of the no. of active (horizontal and vertical) t-bonds .vs. iteration time for a RRTN lattice with $L=80$, $p=0.4$ under external voltage $V=10$. This shows that each kind of the number of active t-bonds asymtotically becomes constant after some time scale (say, $\sim \tau_t$); (b) The exponential tail in the bulk current relaxation of this RRTN sample after a crossover from an non-exponential regime. The behaviour starts about the iteration time of the order of $\tau_t$ and lasts upto the time when the bulk current becomes steady. We call this time as steady-state time, $\tau_s$. }
\label{tstat}
\end{figure}

We observed in general that the RRTN bulk current relaxation dynamics possesses two distinct temporal regimes, i.e., an initial-time non-exponential and a latter-time purely exponential tail with a single time constant. This is obviously an interesting characteristic for this dynamics where exponential regime (signature of a linear-res-ponce behaviour) evolves after a non-exponential/ out-of-linear regime. The relaxation behaviour for any RRN lattice (i.e., a RRTN with no t-bond) is always of a perfect exponential type. Again in a special situation of a RRTN lattice where all of the t-bonds, present there, remain active (i.e., for all t-bonds $|v| \geq v_g$) during the relaxation process, one finds also an exponential dynamics because this will generate practically a modified RRN. Thus the origin of the non-exponential behaviour in the bulk RRTN dynamics is absolutely due to the {\it intermittent} presence (in time) of the active tunneling bonds within RRN skeleton. 
For this reason, we studied the temporal variation in the number of active t-bonds (both horizontal and vertical) with arbitrary microscopic voltage configuration at each node of the RRTN lattice. We denote the total number of active t-bonds at time $t$ as $N_t(t)$. We observed that the number finally saturates to a constant value asymptotically. The behaviours are qualitatively same for all kinds of t-bonds (i.e., horizontal, vertical), so as in the total number of active tunneling bonds. However the number of active vertical bonds are much larger than the active horizontal bonds after same time, supporting that the directed percolation is taking place during RRTN relaxation. In fig. [\ref{tstat}](a), we find this behaviour for a RRTN sample with $L=80, p=0.4$ under an external voltage $V=10$. The (iteration) time since which the total number of t-bonds saturates to a constant value is newly termed here as $\tau_t$. For the sample shown in the fig. [\ref{tstat}](a), it is $\sim 5 \times 10^4$ in unit of arbitrary iteration time. 

%%%%%%%%%%%%%%%%%%%%%%%%%%%%%%%%%%%%%%%%%%%%%%%%%%%%%%%%%%%%%%%%%%%%%%%%%%%%%%%%%%%%%%%%%%%%%%%%%%%%%%%%%%%%%%%%%%%%%%%%%%%%%%%%%%%%%%%%%
%%%%%%%%%%%%%%%%%%%%%%%%%%%%%%%%%%%%%%%%%%%%%%%%%%%%%%%%%%%%%%%%%%%%%%%%%%%%%%%%%%%%%%%%%%%%%%%%%%%%%%%%%%%%%%%%%%%%%%%%%%%%%%%%%%%%%%%%%

\noindent For a {\it quantitative} estimate of $\tau_t$ in a particular RRTN sample, we measure $N_t(t)$ for every iteration-time $t$ and compare this between two consecutive iterations. At early-stage of relaxation, this number changes continuously. However close to $\tau_t$, the $N_t(t)$ does not change for a finite time gap. After the gap we may find a minor change in value. As time progresses, the gap increases
and asymptotically since $\tau_t$, the $N_t(t)$ becomes constant in time. So to ascertain the exact value of $\tau_t$ for each RRTN lattice, we have waited for a safe iteration-time gap. We check that during this wait $N_t(t)$ does not vary any more. If this condition is satisfied upto an iteration-time $t$, then we note this time as $\tau_t$. We have tried with different waiting times and then for the present work, we fix the wait for $200$ lattice-scans, which we find as adequate. One may appreciate that the ratio of this waiting time to the actual $\tau_t$ is quite low even for the smallest sample (e.g, $L=20$), so that such a allowance does not lead to serious error. In fig. [\ref{tstat}](b), we show the exponential tail in the corresponding current relaxation dynamics (actually we plotted here, $(I(t)-I_{steady})$ .vs. $t$, as we did since ref. \cite{sust,epl05}). By $I_{steady}$, we mean the bulk current value of the corresponding RRTN when it becomes steady. Interestingly by comparing the figures [\ref{tstat}](a) and [\ref{tstat}](b), we find that the appearance of the exponential tail in the relaxation and the saturation in the total number of t-bonds are almost simultaneous. So the time $\tau_t$ can be identified as a phenomenological time-scale, present in the RRTN dynamics. 
\noindent Again we may identify here another time-scale $\tau_s$ (we call here as {\it steady state} time), which is the iteration time since when the bulk RRTN current becomes steady value, i.e., $I_{steady}$. The bulk current (in absence of any noise) monotonically reaches to its steady value. After being steady, the bulk current does not change any more. So {\it quantitatively} to confirm that $I(t)$ has reached its steady value, one may compare the $I(t)$ for two consecutive iterations, and refer a time as $\tau_s$ since when there is no temporal change in the bulk current between any pair of consecutive iterations. These phenomenological time-scales like $\tau_t$ and $\tau_s$ actually fix the extents of two prime regimes (e.g., roughly non-exponential regime within time [$0,\tau_t$] and the exponential regime [$\tau_t,\tau_s$] with a cross-over between them) of bulk current relaxation in the RRTN.       

%%%%%%%%%%%%%%%%%%%%%%%%%%%%%%%%%%%%%%%%%%%%%%%%%%%%%%%%%%%%%%%%%%%%%%%%%%%%%%%%%%%%%%%%%%%%%%%%%%%%%%%%%%%%%%%%%%%%%%%%%%%%%%%%%%%%%%%%%
\begin{figure}[htb]
\resizebox*{7cm}{7cm}{\rotatebox{270}{\includegraphics{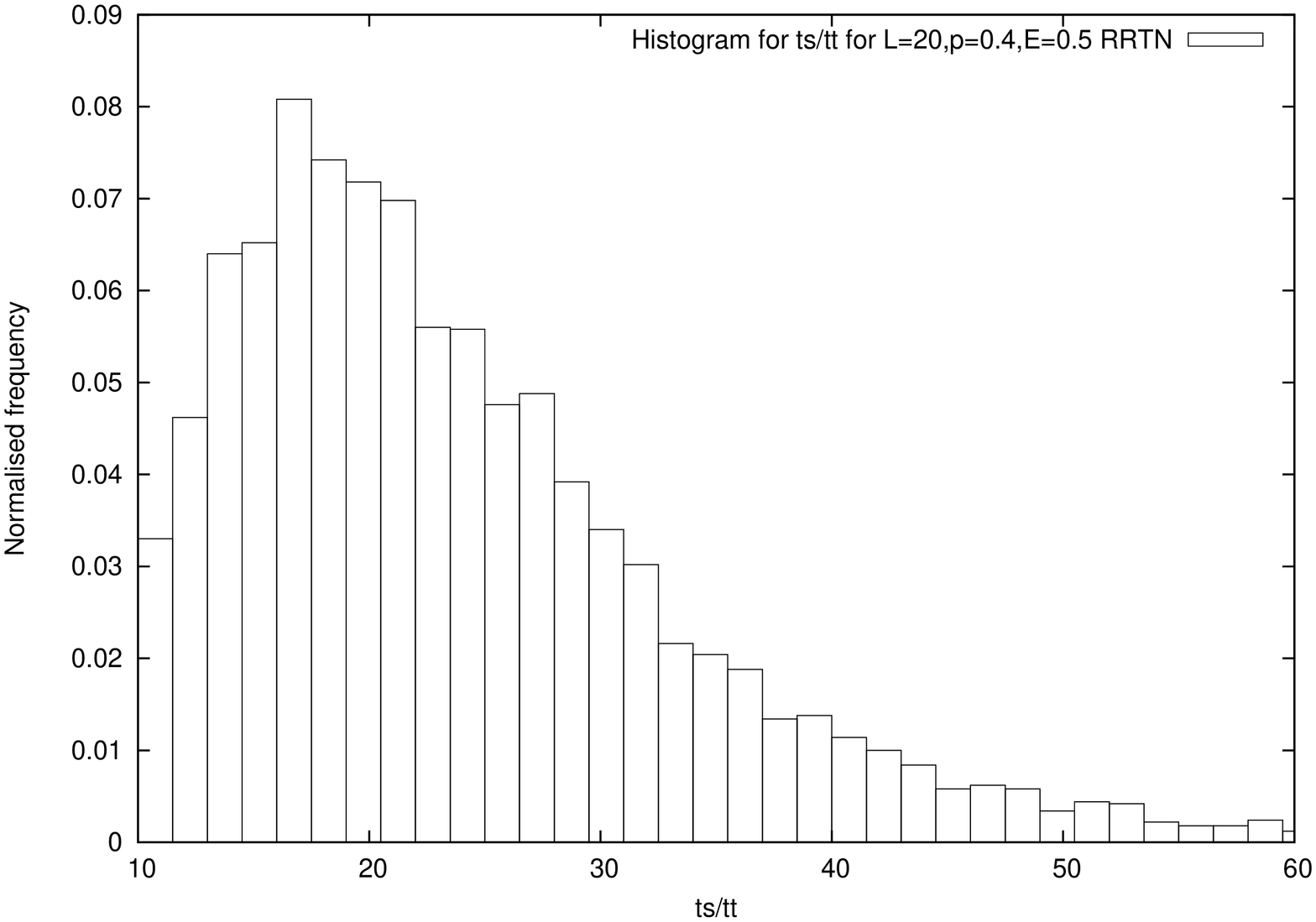}}}
\resizebox*{7cm}{7cm}{\rotatebox{270}{\includegraphics{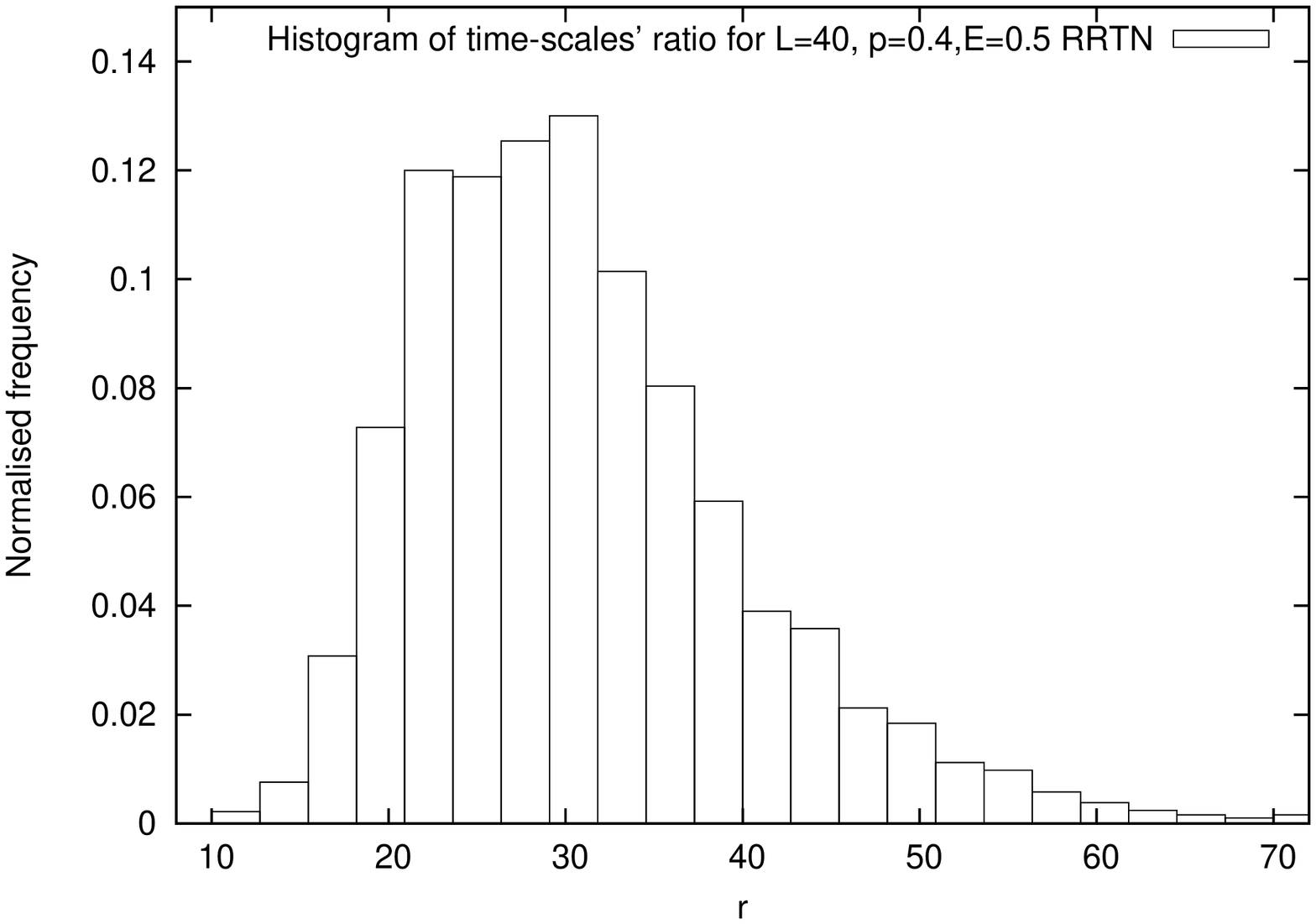}}} 
\caption{(a) The histogram for the ratio of the time-scales (i.e., $\tau_s/\tau_t$) using $N=5000$ random RRTN bond-configurations with system-size $L=20$, ohmic concentration $p=0.4$ for average electric field $E=0.5$. By {\it Normalised frequency}, we mean the actual frequency divided by total number of configurations used i.e., $N$. The behaviour possesses a prominent peak around $16.75$ with a heavy tail; (b) A similar histogram for RRTN samples with system size $L=40$, showing a broad peak. On smoothening we find the mode around $30.25$. The behaviour has relatively weak tail in comparison to (a).}
\label{hisrat1}
\end{figure}
%%%%%%%%%%%%%%%%%%%%%%%%%%%%%%%%%%%%%%%%%%%%%%%%%%%%%%%%%%%%%%%%%%%%%%%%%%%%%%%%%%%%%%%%%%%%%%%%%%%%%%%%%%%%%%%%%%%%%%%%%%%%%%%%%%%%%%%%%
\begin{figure}[htb]
\resizebox*{7cm}{7cm}{\rotatebox{270}{\includegraphics{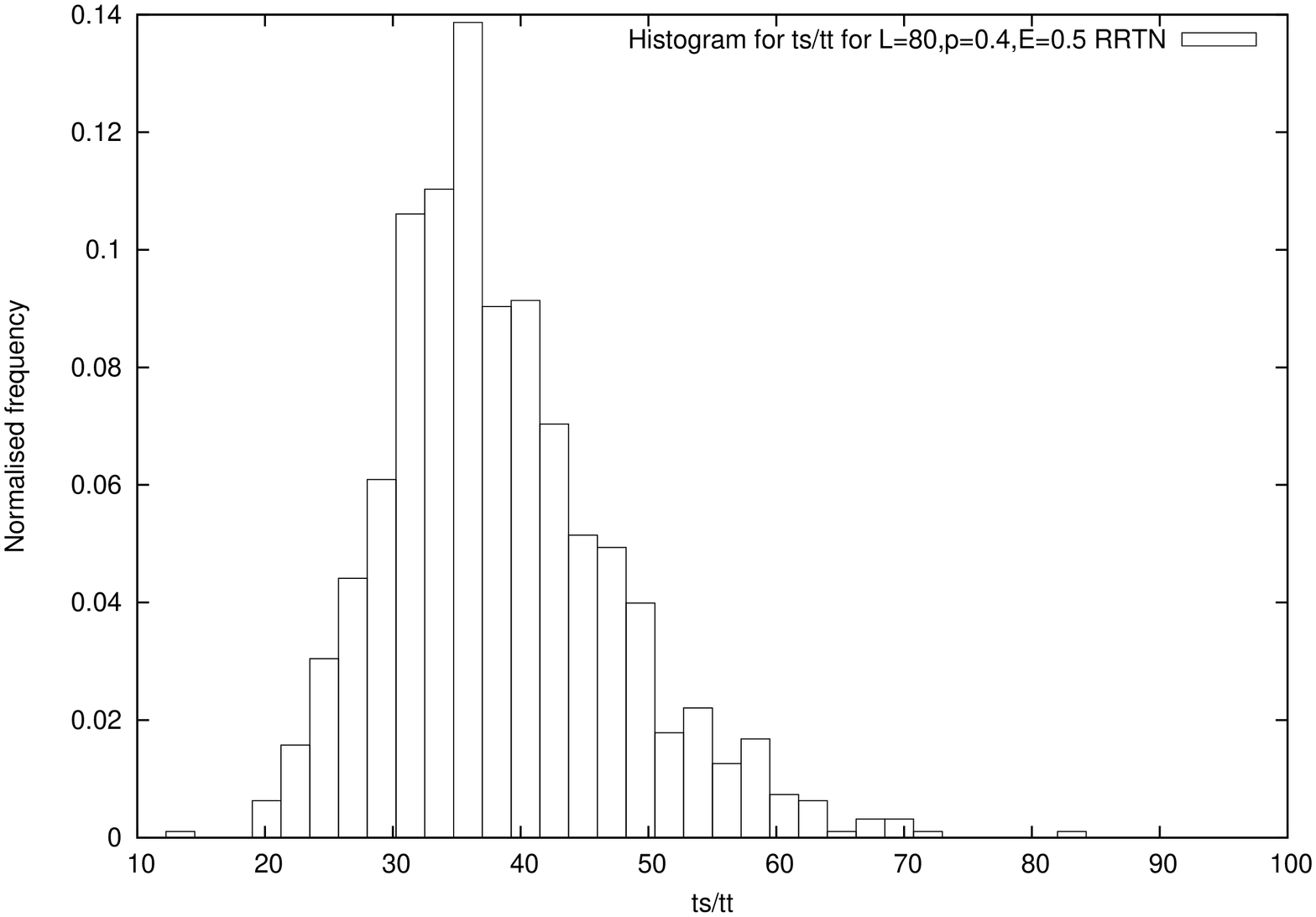}}}
\resizebox*{7cm}{7cm}{\rotatebox{270}{\includegraphics{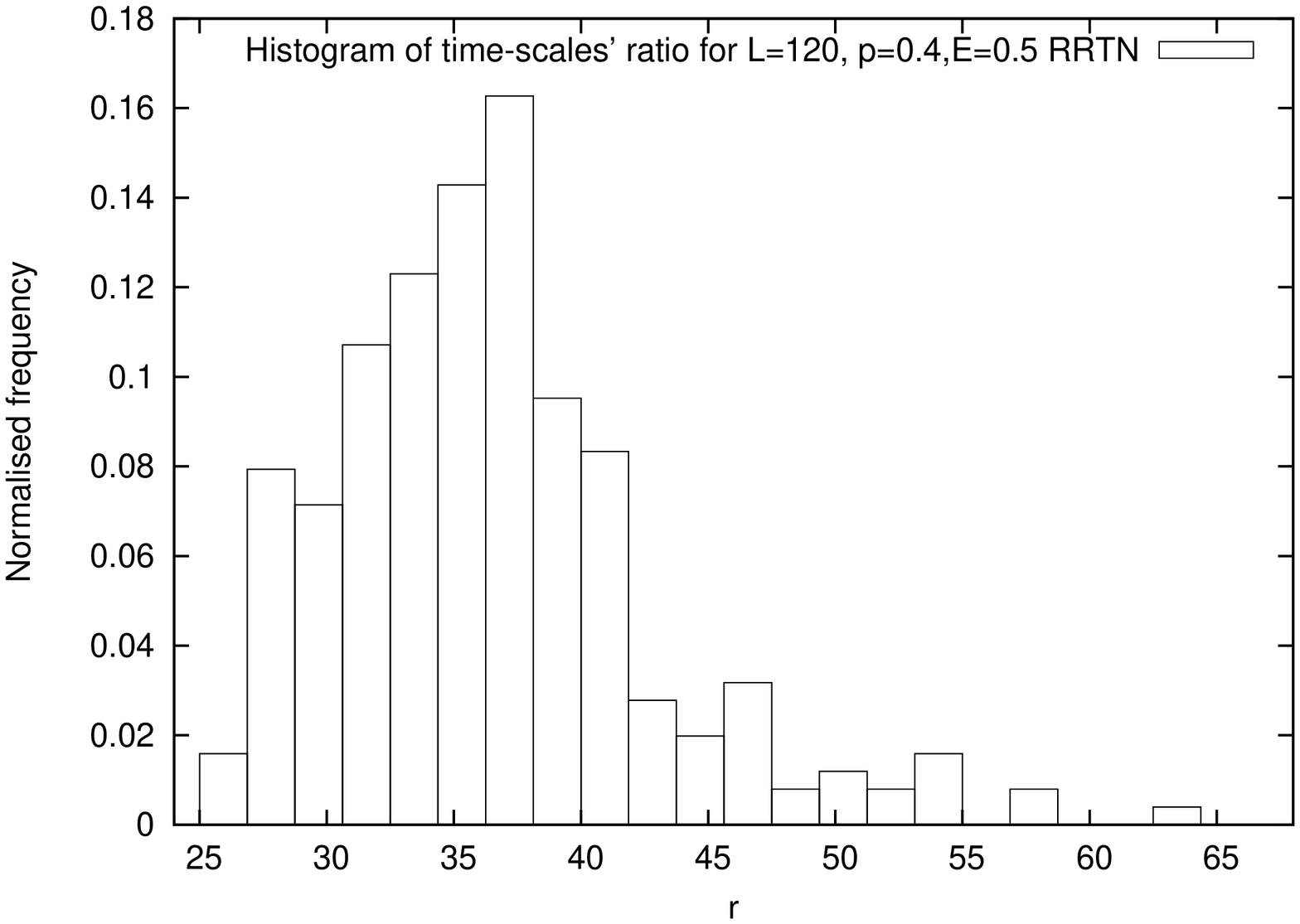}}}
\caption{(a) A histogram for RRTN samples with system size $L=80$, showing mode around $35.88$. There is a sharp peak, almost symmetric around its mode.; (b) A similar histogram for RRTN samples with system size $L=120$, showing a mode around $37.20$. }
\label{hisrat2}
\end{figure}
%%%%%%%%%%%%%%%%%%%%%%%%%%%%%%%%%%%%%%%%%%%%%%%%%%%%%%%%%%%%%%%%%%%%%%%%%%%%%%%%%%%%%%%%%%%%%%%%%%%%%%%%%%%%%%%%%%%%%%%%%%%%%%%%%%%%%%%%%
\subsection{Correlation between the time-scales $\tau_t$ and $\tau_s$}

\begin{figure}[htb]
\centering
\resizebox*{7cm}{7cm}{\rotatebox{270}{\includegraphics{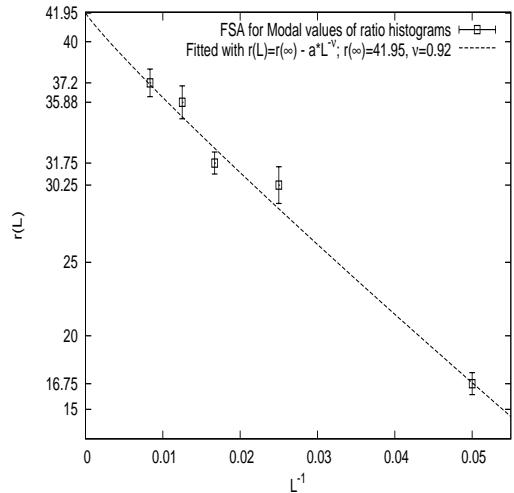}}}
\caption{(a) A graph showing the finite size analysis (FSA) on the mode values of the previous histograms (i.e., Figs. [\ref{hisrat1},\ref{hisrat2}]) using an extrapolation formula, $r(L)=r(\infty)-aL^{-\nu}$. On fitting, we find, $r(\infty) \sim 41.95$ and $\nu =0.92$.  }
\label{fsa}
\end{figure}

To investigate that whether the two time scales $\tau_t$ and $\tau_s$, those evolve from the active t-bond time-series and the bulk current relaxation, are independent or not, we have measured both of them for different RRTN bond-configurations with system sizes 
$L=20 - 120$.  Here we present our work on $5000$ RRTN samples of ohmic concentration $p=0.4$ under external electric field 
$E \equiv \frac{V}{L} =0.5$ with arbitrary initial voltage configuration. The microscopic voltage across each t-bond determines whether it will be active or not. So the number of active t-bonds depends on the electric field (i.e., average voltage per layer). Thus we keep $E$ to be equal for all $L$. In each case (i.e., for $L=20 - 120$), we have plotted the probability density function (or histogram with `normalised frequency') for both $\tau_t$ and $\tau_s$. By {\it normalised frequency}, we mean the actual frequency of occurrence divided by total number of configurations used. All the histograms are observed to have prominent peaks with permissible widths. We identify the mode of each histogram and refer it as $\tau_t(L)$ (in case of $\tau_t$ histogram for system size $L$) or $\tau_s(L)$ (in case of $\tau_s$ histogram for system size $L$). Both $\tau_t(L)$ and $\tau_s(L)$ vary with system size. With this, we also plot the histograms for the ratio of the time-scales, i.e., $r=\tau_s/\tau_t$, for different system sizes and refer the mode value of each histogram as $r(L)$. In figures [\ref{hisrat1},\ref{hisrat2}], we present the histograms for the distribution of $r(L)$ for system sizes $L=20, 40, 80, 120$. The mode of each histogram has been mentioned in the caption of the relevant figures. One may appropriately note here that the mode value of any histogram is inherently associated with an error-bar, which is half of the bin-size of the corresponding histogram. But intriguingly, we find that the $r(L)$s appear to converge to a finite value with increasing system sizes. We use the standard finite size analysis (FSA) to extrapolate the mode value in the thermodynamic limit (i.e., for $L \to \infty$). In fig. [\ref{fsa}], we show our FSA on those $r(L)$s using a conventional extrapolation formula like, $r(L)=r(\infty)-aL^{-\nu}$.  During fitting, we have considered each $r(L)$ with its associated error-bar. We obtain for our work, the parameters values as $r(\infty) = 41.96$ and $\nu = 0.92$. This analysis enables us to comment that though two time-scales like $\tau_t$ and $\tau_s$ separately appear during relaxation dynamics and they quantitatively measure the extents of the RRTN relaxation regimes, but they are purely correlated for the large samples. That is out of those two scales, only one is independent. There is a single time-scale independently present in the RRTN relaxation, that controls the entire bulk current dynamics. 

%%%%%%%%%%%%%%%%%%%%%%%%%%%%%%%%%%%%%%%%%%%%%%%%%%%%%%%%%%%%%%%%%%%%%%%%%%%%%%%%%%%%%%%%%%%%%%%%%%%%%%%%%%%%%%%%%%%%%%%%%%%%%%%%%%%%%%%%%
%%%%%%%%%%%%%%%%%%%%%%%%%%%%%%%%%%%%%%%%%%%%%%%%%%%%%%%%%%%%%%%%%%%%%%%%%%%%%%%%%%%%%%%%%%%%%%%%%%%%%%%%%%%%%%%%%%%%%%%%%%%%%%%%%%%%%%%%%

\section{Conclusion}

In several research works (experiment as well as theory) on diverse type of disordered systems, the origin of non-exponential relaxation behaviour is identified with local structural rearrangements followed by the final global structural rearrangements among the components.
For RRTN, the basic structure of the bond-network in a particular sample is created once for all and it is the microscopic voltages across the bonds which keep changing during the entire dynamics. This generates two distinct microscopic mechanisms within RRTN, which are behind
the far-from-steady state relaxation behaviour in the bulk current. In RRN, only the {\it linear} resistors (i.e., o-bonds) are conducting currents, so that the microscopic voltages at all the nodes are being updated equivalently. The insulating bonds are always inactive in RRN. As the transport through the o-bonds are purely diffusive, so the current relaxation in the RRN belongs to Debye-class.  

\noindent However for RRTN, in addition to o-bonds, some of the insulators were been upgraded as t-bonds, which behave {\it nonlinearly} under field. The behaviour is nonlinear as the activation of a t-bond depends on the voltage across it. During relaxation process as the voltage at each node changes for different times, so the activation of a t-bond becomes a highly time-dependent phenomenon. At every node for the RRTN more than one bonds meet. For example, it is four for inside, three for edges and two at the corners. If atleast one t-bond meets at any node, the updating of the microscopic voltage at that node becomes qualitatively different than the similar mechanism that runs for a linear network like RRN. Here it updates due to a mechanism as in a linear network (where the microscopic conductance of a bond is time-independent), which is significantly modulated by the intermittent presence of participating t-bonds. So in the microscopic length-scale, two different kinds of mechanisms (based on Gauss-Seidel algorithm) are running simultaneously inside RRTN during the early-stage non-exponential relaxation. The frequent change in the participation from the nonlinear t-bonds is responsible for the non-Debye type of behaviour in the RRTN current relaxation during the early-time. In general any scale-free behaviour is expected only in a finite time domain, after which it is supposed to crossover to the bulk dynamics. Similarly, the nonlinear process stops around the time-scale $\tau_t$, when no further t-bond becomes active/de-active thereafter (See fig. [\ref{tstat}](a)). So the time-dependent appearance of tunneling bonds, which were influencing the current relaxation previously cease to affect. Rather now, all the active t-bonds contribute constantly in time till the bulk current becomes steady (around $\tau_s$). This situation is qualitatively identical with that during the current relaxation in the RRN, as then a RRTN roughly becomes an effective RRN with mostly two kinds of linear bonds. For this, we observe an asymptotic purely exponential tail (See fig. [\ref{tstat}](b)). The extent of these two qualitatively different behaviours in time can be understood by looking at the time-scales like $\tau_t$ and $\tau_s$. We find that though they are present in every RRTN relaxation behaviour but their values are perfectly correlated. So that measurement of one enables us to predict the other. This means out of two, only one of them is independent. In that sense, there exists a single time-scale in the RRTN current relaxation process. 

\noindent In this same paper, we also discuss on another interesting feature of the RRTN dynamics, the strong convergence of bulk current towards its robust steady-state. We propose an argument behind this from the general properties of Gauss-Seidel relaxation method, for solving a set of coupled algebraic equations.

\section*{Acknowledgements}
The author admits some useful discussions during the presentation of the work in several conferences.
He also acknowledges the full support by UGC Minor Research Grant File No. {\it PSW-162/14-15 (ERO) 
dt. 3.2.15} in completion and presentation of the work. Author remembers some helpful suggestions from 
Mr. Sudip Mukherjee, Barasat Govt. College, during preparation of the manuscript.

%\section*{References}

\end{document}